# Latent and sensible heat fluxes overestimated and net heat flux underestimated in Lake Victoria

Piet Verburg
National Institute of Water and Atmospheric Research, Hamilton, New Zealand

Cozar et al. (2012) used remotely-sensed data to link phytoplankton growth to water column stability. The authors inferred a lake-wide convective circulation from differential cooling between the northern and southern parts of the lake, which was concluded to affect phytoplankton growth. The differential cooling was inferred from surface temperatures and the heat balance where the net heat flux was estimated as the difference between the gains of heat by solar and incoming long wave radiation and losses by outgoing long wave radiation, latent and sensible heat fluxes. The phytoplankton biomass was found to be coupled to the net heat flux in both the northern and southern parts of the lake, while the lake surface temperature did not show the seasonal north-south shift apparent in phytoplankton biomass.

The latent and sensible heat fluxes are proportional to air density, as shown in the equations of Cozar et al. (2012). Cozar et al. (2012) assumed a constant air density of 1.3 kg m$^{-3}$ in the calculation of latent and sensible heat fluxes. This value for mean air density at Lake Victoria is much too high. Even at sea level (average air pressure 1013 mb) freezing air temperatures would be needed to find such a high air density. Air density is not constant but varies with latitude, altitude, seasons and time of day. Air density decreases with decreasing air pressure, with increasing air temperature and with increasing relative humidity. As a result, air density is generally lower in the tropics and at high altitudes such as at Lake Victoria, compared with sea level locations in temperate regions.

The equation to estimate air density (kg m$^{-3}$) is given (Chow et al. 1988; Verburg and Antenucci 2010) by

$$\rho_a = 100p/[R_a(T+273.16)]$$

where T is the air temperature (°C), p is the air pressure (mb) and $R_a$ is the gas constant for moist air (J kg$^{-1}$ K$^{-1}$)

$$R_a = 287(1 + 0.608q_z)$$

$q_z$ is the specific humidity (kg kg$^{-1}$)

$$q_z = 0.622e_a/p$$

$e_a$ is the vapor pressure (mb)

$$e_a = R_H e_s/100$$

$R_H$ is the relative humidity (%) and $e_s$ is the saturated vapor pressure at T (mb)

$$e_s = 6.11\exp^{[17.27T/(237.3 + T)]}.$$

Lake Victoria lies at an altitude of 1133 m, and has a mean atmospheric pressure equal to about 885 mb (http://www.windfinder.com/forecast/entebbe). The average air temperature at stations around the lake is about 22.3 °C (Yin and Nicholson 1998) although above the lake surface it is likely to be higher, because the monthly mean temperature of the water surface ranges between 24.5 and 27.5°C (Cozar et al. 2012; Yin and Nicholson 1998; MacIntyre et al. 2002). The mean relative humidity is about 70% (MacIntyre et al. 2002; http://rp5.ru/Weather_in_Entebbe_(airport) ). Of the three variables needed to estimate air density, air pressure has the largest effect and relative humidity the least.

From p = 885 mb, T = 22.3 °C, and $R_H$ = 70%, it follows that $\rho_a$ = 1.035 kg m$^{-3}$. That means that the latent and sensible heat fluxes were overestimated by 26% by Cozar et al.

(2012). With a higher mean air temperature, T = 25 °C, $\rho_a$ = 1.024 kg m$^{-3}$, indicating overestimation by 27%. As a result of overestimating the latent and sensible heat fluxes, the net heat flux of Cozar et al. (2012), estimated as the difference between inputs and outputs of heat from the lake, was severely underestimated. After correction for the overestimation of the latent and sensible fluxes, the net heat fluxes in the southern and northern regions of the lake are roughly +15 W/m2 and +50 W/m2, respectively, no longer of opposite sign. This brings into question the conclusions regarding the convective circulation. However, the consistent surface temperature difference between the northern and southern parts of the lake and the low values of the dimensionless shear parameter B (Cozar et al. 2012) suggest other sources of error in the methods of Cozar et al. (2012) may have made up for the difference resulting from the error in the estimation of air density.